\documentclass{Rinton-P10x7}

\begin{document}



\title{Issues on Radiatively Induced Lorentz and CPT Violation
in Quantum Electrodynamics }

\author{ W.F. Chen }

\address{Department of Mathematics and Statistics, University of
Guelph\\ Guelph, Ontario, Canada N1G 2W1 \\
E-mail: wchen@uoguelph.ca}

\maketitle

\abstracts{Various ambiguous results on radiatively induced
Lorentz and CPT violation  in quantum electrodynamics with a
modified fermionic sector are reviewed and  possible explanations
for this ambiguity appearing in the literature are commentated.
Furthermore, joint between stringent limit from astrophysical
observation and theoretical prediction on Lorenz and CPT violation
is discussed. }


\section{Introduction}

Lorentz symmetry is algebraic foundation of the theory of special
relativity. Nearly one hundred years the theory of special
relativity keeps the status as a cornerstone of modern physics and
has been supported by numerous high energy physics experiments and
astrophysical observation. However, physics is a science born out
of experimental observation. With the availability of higher
precision experimental or observational data, it is conceivable
that even the most fundamental  principles may someday have to be
modified or even abandoned. There are quite a number of such
examples in the history of physics. It is partly in this spirit
that an investigation on the possible breaking of Lorentz symmetry
is not fantastic.

In fact, the spontaneous breaking of Lorentz symmetry is a natural
consequence of string theory. If the Standard Model is considered
as the low-energy limit of a more fundamental theory constructed
from string, the spontaneous breaking of Lorentz symmetry can
occur naturally since string theory generally involves
interactions that make a Lorentz tensor get non-zero vacuum
expectation value\cite{kosa}.

A straightforward reason of considering  Lorentz and CPT violation
in quantum electrodynamics (QED) was from astrophysical
observation. A lopsided analysis on the polarized electromagnetic
radiation emitted by distant radio galaxies revealed that the
universe may present cosmological anisotropy in electromagnetic
wave propagation\cite{nora}. Moreover, the analysis suggested that
this chiral effect can be well described at lower derivative
expansion by a modified classical electrodynamics proposed a
decade ago\cite{cfj},
\begin{eqnarray}
S = \int d^4x\left(-\frac{1}{4}F_{\mu\nu}F^{\mu\nu} +\frac{1}{2}
\epsilon^{\mu\nu\lambda\rho}k_\mu F_{\nu\lambda}A_\rho\right).
\label{eq:med}
\end{eqnarray}
The first term of (\ref{eq:med}) is the familiar Maxwell term, and
the second one is called the Chern-Simons-like (CS) term,
\begin{eqnarray}
{\cal L}_{\rm CS}=\frac{1}{2}\epsilon^{\mu\nu\lambda\rho} k_\mu
F_{\nu\lambda} A_\rho, \label{csterm}
\end{eqnarray}
which explicitly violate Lorentz and discrete CPT symmetries since
 $k_\mu$ is certain background constant vector
in four-dimensional space-time. Despite that a more rigorous
analysis on the astrophysical observation data has excluded the
polarization effect of electromagnetic wave in propagating from
the distant radio sources\cite{cafi}, this still provides a
promising way to observe a violation of Lorentz and CPT symmetries
in nature, and hence stimulates an effort to explore a possible
Lorentz and CPT violating mechanism theoretically.

Furthermore, a $SU(3)\times SU(2)\times U(1)$ standard model with
explicit Lorentz and CPT violating extension had been constructed
and hence a quantitative physical theory of studying Lorentz and
CPT violation was furnished\cite{coko}. Predictions on the
possible Lorentz and CPT violation from this extended Standard
Model can be tested by high-precision measurements in numerous
existing experiments and possibly in next generation
accelerator\cite{kos}.

In this talk I shall concentrate on a typical quantum field theory
problem, namely, whether the CPT-odd pure photon term in
electrodynamics, i.e. the CS term shown in Eq.\,(\ref{csterm}),
can be induced from quantum correction with a modified fermionic
sector
\begin{eqnarray}
{\cal L}_{\rm fermion}
=\overline{\psi}\left(i\partial\hspace{-2mm}/\,-eA\hspace{-2mm}/\,
-b\hspace{-1.5mm}/\,\gamma_5-m\right)\psi, \label{fermion1}
\end{eqnarray}
where  $b_\mu$ is a constant prescribed four-vector. The new
introduced gauge invariant interaction term between constant
vector $b_\mu$ and axial vector current $j_\mu^5
(x)=\overline{\psi}\gamma_\mu\gamma_5\psi$ violates Lorenz and CPT
symmetries explicitly, since $b_\mu$ picks up a fixed direction in
space-time. If the CS term can  be induced from the radiative
correction with the coefficient $k_\mu\propto b_\mu$, then a
constraint on ${\cal L}_{CS}$ from astrophysical observation will
restrict a possible Lorentz and CPT violation in the fermionic
sector.

 In section 2 we shall review various results on radiatively induced
 CS term, and then  analyze the
 possible origin for this ambiguity in section 3. Finally we summarize
  and discuss the joint of astrophysical observation with theoretical
  prediction on Lorentz and CPT violation and some relevant problems.

\section{Various Controversial and Ambiguous Results on Radiatively
Induced Chern-Simons Term}

As a general procedure, the quantum effective action can be
obtained by integrating out fermionic fields,
\begin{eqnarray}
e^{i\Gamma[A,b]}&=&\int D\overline{\psi}D\psi e^{i\int d^4x
\overline{\psi}\left(i\partial\hspace{-1.7mm}/\,-eA\hspace{-1.8mm}/\,
-b\hspace{-1.4mm}/\,\gamma_5-m\right)\psi}\nonumber\\
&=&\det\left(i\partial\hspace{-2mm}/\,-eA\hspace{-2mm}/\,
-b\hspace{-1.6mm}/\,\gamma_5-m\right);\nonumber\\
 \Gamma[A,b] &=& -i\mbox{Tr}\ln
\left(i\partial\hspace{-2mm}/\,-eA\hspace{-2mm}/\,
-b\hspace{-1.6mm}/\,\gamma_5-m\right). \label{eq:efac}
\end{eqnarray}
The radiatively induced Chern-Simons term will be $b$-linear and
parity-odd part of above effective action.

It is well known that $\Gamma [A,b]$ or equivalently the relevant
fermionic determinant cannot be evaluated exactly. A perturbative
expansion or certain approximation must be utilized and a use of
 regularization scheme must be made in the calculation.
 At first sight, the evaluation of CS term in the
 quantum effective action is a typical and simple quantum field
 theory problem. However, the concrete calculation turned out to
 be rather non-trivial and the result presented remarkable
 ambiguities: distinct relations between $k_\mu$ and $b_\mu$ can
 yield depending on concrete calculation schemes.
 The various ambiguous results are listed  in the following:
 \begin{itemize}
\item \begin{eqnarray}
    k_\mu=0.
\end{eqnarray}
Coleman and Glashow argued first that the CS term cannot be
generated\cite{cogl}. They considered that the axial vector
current $j_\mu^5(x)=\overline{\psi}(x)\gamma_\mu\gamma_5\psi(x)$
should keep gauge invariant in the quantum theory at any momentum
or equivalently at any space-time point. Since $\langle j_\mu^5
(x)\rangle =\delta {\cal L}(x)/\delta b_\mu$, this hypothesis is
actually equivalent to the requirement that the Lagrangian density
corresponding to the quantum effective action should be gauge
invariant. Thus, based on this requirement, the CS term cannot be
generated since its Lagrangian density is explicitly not invariant
under gauge transformation $A_\mu\rightarrow A_\mu+\partial_\mu
\Lambda$. Furthermore, Bonneau studied the renormalization of an
extended QED including the CS term (\ref{csterm}) and the modified
fermionic sector (\ref{fermion1}). He found that Ward identities
and the renormalization conditions  determine uniquely the absence
of CS term from quantum correction\cite{bo}. In fact, when
deriving Ward identities, Bonneau introduced external source
fields for the axial vector current and the CS term, so the Ward
identities he derived actually impose gauge invariance on
Lagrangian density and hence coincide with the ``no-go theorem"
requirement argued by Coleman and Glashow\cite{cogl}.  Recently,
Adam and Klinkhamer put forward an independent line of reasoning
the vanishing of radiatively induced CS term rather than gauge
symmetry\cite{adkl}. They argued that if the perturbative
expansion near $b^2=0$ is valid and further the gauge field
$A_\mu$ is regarded as a quantized dynamical field rather than an
external background field as Bonneau did, the presence of CS term
with a purely time-like coefficient may violate the causality
principle of a quantum field theory. This argument thus excluded
the radiative induction of CS term. In addition, explicit
perturbative calculations in Pauli-Villars
regularization\cite{coko} and dimensional regularization\cite{bo}
shown that $k_\mu$ should vanish.
\item
\begin{eqnarray}
k_\mu=\frac{3e^2}{16\pi^2}b_\mu. \label{eq:re2}
\end{eqnarray}
However, Jackiw and Kosteleck\'{y} thought that since $j_\mu^5(x)$
only couples with a constant 4-vector $b_\mu$, it is true to
require only that $j_\mu (x)$ with zero-momentum (i.e. $\int
d^4xj_\mu^5 (x)$) is gauge invariant at quantum level. Since
$\langle\int d^4xj_\mu^5 (x)\rangle =\delta S/\delta b_\mu$, this
statement is equivalent to the requirement that the quantum
effective action should be gauge invariant. Thus the dynamical
generation of CS term can escape from above ``no-go theorem"
conjectured by Coleman and Glashow, since the action of CS term is
gauge invariant. Based on this observation, Jackiw and
Kosteleck\'{y} calculated the $b$-linear part of the one-loop
vacuum polarization tensor with $b_\mu$-exact propagator. They
ingeniously manipulated the linear divergent term in the loop
integration by writing it into a finite term plus an external
momentum derivative term, and then arrived at the finite result
(\ref{eq:re2}). Actually, this result was obtained earlier by
Chung and Oh in calculating the one-loop effective action
Eq.\,(\ref{eq:efac}) in terms of dimensional regularizatoion plus
derivative expansion\cite{choh}. Furthermore,
P\'{e}rez-Victoria\cite{victoria1} and Chung\cite{chung1} proved
through an explicit calculation on the parity-odd part of one-loop
vacuum polarization tensor at zero external momentum that the
result (\ref{eq:re2}) stands to any order of $b_\mu$.
\item
\begin{eqnarray}
k_\mu=C b_\mu, ~~~\mbox{$C$ being an arbitrary constant}.
\label{eq:re3}
\end{eqnarray}
It was first realized by Jackiw that the  perturbative ambiguity
for radiatively induced CS term can be revealed quantitatively in
a new developed regularization method called differential
regularization\cite{djl}. This new calculation technique works for
a Euclidean field theory in coordinate space. Its invention is
based on the observation that the short-distance singularity
representing the UV divergence of a primitively divergent Feynman
diagram prevent the amplitude from having a Fourier transform into
momentum space. So one can regulate such an amplitude by writing
its singular term as a derivative of another function, which has a
well defined Fourier transform, then performing Fourier transform
into momentum space through partial integration and discarding the
surface term. In this way one can directly arrive at a
renormalized amplitude. The great advantages of this
regularization over conventional regularization methods lie in
that it does not modify the original classical action and in
particular, it does not impose or violate gauge symmetry in the
process of calculation. Only at the end of calculation, one can
get the preferred symmetry by an appropriate choice on indefinite
renormalization scales. This is the reason why differential
regularization can  clearly show the perturbative ambiguity. The
$b$-linear part of one-loop vacuum polarization tensor in
differential regularization reads\cite{chen}
\begin{eqnarray}
 \Pi^{(b)}_{\mu\nu}(x)
&=&\frac{m^2}{2\pi^4}b_{\lambda}\epsilon_{\lambda\mu\nu\rho}\left\{
-2\left[\frac{\partial}{\partial
x_\rho}\left(\frac{K_1(mx)}{x}\right)^2 - \frac{\partial}{\partial
x_\rho}\left(\frac{K_1(mx)}{x}\right)^2\right]
\right.\nonumber\\
&&\left. +m\frac{\partial}{\partial
x_\rho}\left[\frac{K_1(mx)K_0(mx)}{x}\right]\right\}\nonumber\\
&=&\frac{m^2}{2\pi^4}b_{\lambda}\epsilon_{\lambda\mu\nu\rho}
\left\{-\frac{1}{2m^2}\frac{\partial}{\partial x_a}\left[
\partial^2\left(\frac{\ln x^2M^2_1}{x^2}\right)- \partial^2
\left(\frac{\ln x^2M^2_2}{x^2}\right)\right]
\right.\nonumber\\
&&\left. +m\frac{\partial}{\partial x_\rho}
\left[\frac{K_1(mx)K_0(mx)}{x}\right]\right\}\nonumber\\
&=&\frac{1}{2\pi^4}b_{\lambda}\epsilon_{\lambda\mu\nu a}
\left\{4\pi^2 \ln\frac{M_1}{M_2}\,\frac{\partial}{\partial
x_\rho}\delta^{(4)}(x)
+m^3\frac{\partial}{\partial x_\rho}
\left[\frac{K_1(mx)K_0(mx)}{x}\right]\right\},
\end{eqnarray}
where we have used the asymptotic expansion of the first-order
modified Bessel function of the second kind near $x^2\sim 0$,
\begin{eqnarray}
\frac{K_1(mx)}{x}\stackrel{x^2\sim 0}{\longrightarrow}
\frac{1}{mx^2}+\frac{1}{4} m\ln (m^2 x^2)+\cdots,
\end{eqnarray}
and the fact that the UV divergence is only contained in the
leading term  as well as a typical differential regularization
identity,
\begin{eqnarray}
\left(\frac{1}{x^4}\right)_{R}=-\frac{1}{4}\partial^2\left[\frac{\ln
(x^2 M^2)}{x^2}\right].
\end{eqnarray}
The corresponding Fourier transform is thus
\begin{eqnarray}
\Pi^{(b)}_{\mu\nu}(p) &=& \int d^4xe^{-ip{\cdot}x}\Pi_{\mu\nu}(x)
=\frac{2}{\pi^2}b_{\lambda}\epsilon_{\lambda\mu\nu\rho}ip_\rho
\left[\ln\frac{M_1}{M_2}
\right.\nonumber\\
&&\left. +\frac{m}{4p\sqrt{1+p^2/(4m^2)}}
\ln\frac{\sqrt{1+p^2/(4m^2)}+p/(2m)}{\sqrt{1+p^2/(4m^2)}-p/(2m)}\right],
\end{eqnarray}
and the CS term is relevant to above polarization tensor at
low-energy limit,
\begin{eqnarray}
\left.\Pi^{(b)}_{\mu\nu}(p)\right|_{p^2=0}=\frac{2i}{\pi^2}\epsilon_{\rho\mu\nu
\lambda} b_\rho p_\lambda
\left(\ln\frac{M_1}{M_2}+\frac{1}{4}\right). \label{eq25}
\end{eqnarray}
Since $M_1$ and $M_2$ are two arbitrary renormalization scales,
the above result shows that the relation between $k_\mu$ and
$b_\mu$ is completely arbitrary.

 Furthermore, the same conclusion was drawn by Chung\cite{chung2}
through an analysis on the non-invariance of path integral measure
under axial vector gauge transformation as Fujikawa's method of
evaluating chiral anomaly, and then an explicit calculation on the
$b$-linear part of the vacuum polarization tensor with the action
manifesting the non-invariance under axial vector gauge
transformation.

\item
\begin{eqnarray}
k_\mu=\frac{e^2}{8\pi^2}b_\mu \label{eq:re4}
\end{eqnarray}
It was found by Chan\cite{chan}, when adopting the covariant
derivative expansion\cite{chga} to evaluate the anomalous
contribution to the effective action (\ref{eq:efac}), that due to
the noncommutativity of the operators $\partial$ and $A_\mu(x)$
there arises a non-Feynman diagram contribution to the $b$-linear
part of the vacuum polarization tensor. This additional term looks
quite exotic from the viewpoint of perturbation theory and it
seems to represent a somehow non-perturbative contribution. As a
consequence, the result (\ref{eq:re2}) was modified to
$e^2/(8\pi^2)b_\mu$. The basic idea of covariant derivative
expansion is to express  local quantum effective Lagrangian in
powers of gauge covariant derivative $\Pi_\mu=
i\partial_\mu-eA_\mu$ rather than in powers of $i\partial_\mu$ and
$A_\mu$ separately\cite{chga}. The remarkable difference between
(\ref{eq:re2}) and (\ref{eq:re4}) and the feature of covariant
derivative expansion motivated us to re-calculate the $b$-linear
part of the effective action (\ref{eq:efac}) in the technique of
Schwinger's constant field approximation\cite{sch} since this
method shares the same feature as covariant derivative expansion.
The essence of this method is converting the calculation of an
effective action as (\ref{eq:efac}) into solving a harmonic
oscillator problem in non-relativistic quantum mechanics. It is
worth to mention that this is an evergreen method and has been
utilized in many aspects such as computation of gravitational
anomaly in $4n+2$ dimensions\cite{alwi}, investigation on the
dynamical violation of parity and gauge symmetries in three
dimensions\cite{red} and determining the low-energy effective
action of $N=2$ supersymmetric Yang-Mills theory\cite{ccm} etc. We
found that if the following trace condition is satisfied,
\begin{eqnarray}
\lim_{x^\prime \rightarrow
x}\frac{(x-x^\prime)_{\mu}(x-x^\prime)_{\nu}}{(x-x^\prime)^2} =
\lim_{x^\prime \rightarrow x}\frac{1}{\int d^4x^\prime}\int
d^4x^\prime
\frac{(x-x^\prime)_{\mu}(x-x^\prime)_{\nu}}{(x-x^\prime)^2} =
\frac{1}{4}g_{\mu\nu}, \label{eq:trace}
\end{eqnarray}
then the same result as in covariant derivative expansion can be
reproduced\cite{ccgf},
\begin{eqnarray}
 \left\langle J_{\mu}(x)\right\rangle &=&\frac{\delta \Gamma_{\rm
CS}}{\delta A_{\mu}(x)} = \frac{e^2}{4\pi^2}\left\{
\exp\left[-ie\int_{x^\prime}^x dy^\rho A_{\rho}(y)\right]
\frac{mK_1([-m^2 (x-x^{\prime})^2]^{1/2}}{[-m^2
(x-x^{\prime})^2]^{1/2}}\right.
\nonumber\\
&& \times \left.\left.(x-x^{\prime})_\mu (x-x^{\prime})_\rho
b_{\nu}\epsilon^{\rho\nu\alpha\beta}
F_{\alpha\beta}\right\}\right|_{x^\prime\rightarrow x}\nonumber\\
&=&-\frac{e^2}{16\pi^2}\epsilon^{\mu\nu\lambda\rho}b_\nu
F_{\lambda\rho},\\
\Gamma_{\rm CS}&=&\frac{e^2}{16\pi^2}\int d^4x
\,\epsilon^{\mu\nu\lambda\rho} b_\mu A_\nu F_{\lambda\rho}.
\end{eqnarray}
However, it should be emphasized that the limit given in
Eq.\,(\ref{eq:trace}) has a potential ambiguity and the general
result will be that $\lim_{x\rightarrow 0} x_\mu x_\nu/x^2=C
g_{\mu\nu}$, thus the induced CS term is actually ambiguous.
Furthermore, this ambiguous result was confirmed by Chungs to any
order of $b_\mu$ in the same method\cite{chch}.
 \end{itemize}

\section{Possible Origin of Ambiguity}

Four finite but entirely different  results on radiatively induced
CS term are shown in last section. The reason behind this
ambiguity should be unearthed. Two convincing explanations has
been proposed by Sitenko\cite{sitenko} and
P\'{e}rez-Victoria\cite{victoria2}. The former emphasized the
calculation technique cause, while the latter indicated a
theoretical origin.

Let us first look at the explanation proposed by
Sitenko\cite{sitenko}, which concerns with two different
formulations of the quantum effective action (\ref{eq:efac})
adopted by Chaichian et al\cite{ccm} and Chung et al\cite{choh},
\begin{eqnarray}
\Gamma^{(1)}&=&-i\mbox{Tr}\ln
\left(i\partial\hspace{-2mm}/\,-A\hspace{-2mm}/\,-m\right)
+i\int_0^1dz\,\mbox{Tr}\left[\left(i\partial\hspace{-2mm}/\,
-A\hspace{-2mm}/\,-z\gamma^5
b\hspace{-1.6mm}/-m\right)^{-1}\gamma^5
b\hspace{-1.6mm}/\right],\label{form1}\\
 \Gamma^{(2)}&=&-i\mbox{Tr}\ln
\left(i\partial\hspace{-2mm}/\,-\gamma^5
b\hspace{-1.6mm}/\,-m\right) +i
\int_0^1dz\,\mbox{Tr}\left[\left(i\partial\hspace{-2mm}/\,
-zA\hspace{-2mm}/\,-\gamma^5 b\hspace{-1.6mm}/-m\right)^{-1}
A\hspace{-2mm}/\right]. \label{form2}
\end{eqnarray}
Their $b$-linear sectors read
\begin{eqnarray}
&& i \int_0^1dz\,\mbox{Tr}\left[\left(i\partial\hspace{-2mm}/\,
-A\hspace{-2mm}/\,-z\gamma^5
b\hspace{-1.6mm}/-m\right)^{-1}\gamma^5
b\hspace{-1.6mm}/\right]_{(b)} =\frac{1}{16\pi^2}b_\mu \int d^4x
\epsilon^{\mu\nu\lambda\rho} F_{\nu\lambda}A_\rho\nonumber\\
&+& \frac{1}{8\pi^2}b_\mu I_\nu\int
d^4x\epsilon^{\mu\nu\lambda\rho} F_{\lambda\rho}
+\frac{1}{8\pi^2}I_{\alpha\beta}\left(
g^{\beta\nu}\epsilon^{\mu\alpha\rho\lambda}
+g^{\beta\lambda}\epsilon^{\mu\nu\rho\alpha}\right)b_\mu \int d^4x
A_\nu \partial_\rho A_\lambda, \\
 && i
\int_0^1dz\,\mbox{Tr}\left[\left(i\partial\hspace{-2mm}/\,
-zA\hspace{-2mm}/\,-\gamma^5 b\hspace{-1.6mm}/-m\right)^{-1}
A\hspace{-2mm}/\right]_{(b)} = \frac{3}{32\pi^2}b_\mu \int d^4x
\epsilon^{\mu\nu\lambda\rho}
 F_{\nu\lambda}A_\rho\nonumber\\
&+&\frac{1}{8\pi^2}I_{\alpha\beta} \left(
g^{\beta\nu}\epsilon^{\mu\alpha\rho\lambda}
+g^{\beta\lambda}\epsilon^{\mu\nu\rho\alpha}
+g^{\beta\rho}\epsilon^{\mu\nu\rho\alpha}\right)b_\mu \int d^4x
A_\nu \partial_\rho A_\lambda ,
\end{eqnarray}
where
\begin{eqnarray}
I_\mu &{\equiv}& \frac{1}{i\pi^2}\int
d^4k\frac{k_\mu}{(k^2-m^2)^2},
\nonumber\\
I_{\mu\nu}&{\equiv}&\frac{1}{i\pi^2}\int d^4k\frac{4k_\mu
k_\nu-k^2 g_{\mu\nu}}{(k^2-m^2)^3}
=\frac{1}{2}g_{\mu\nu}-\frac{1}{i\pi^2}\int
d^4k\frac{\partial}{\partial k^\mu}
\left[\frac{k_{\nu}}{(k^2-m^2)^2}\right]
\end{eqnarray}
are two momentum integrals with superficially linear  and
logarithmic divergence, respectively. Due to the explicit breaking
of Lorentz symmetry, one cannot naively put $I_\mu=0$ and use the
formula $\int d^nk k_\mu k_\nu f(k^2)=1/n\,g_{\mu\nu}\int d^nk k^2
f(k^2)$. The approaches of regularization schemes manipulating
these two divergent integrals lead to above finite ambiguities on
the radiatively induced CS term\cite{sitenko}:
\begin{itemize}
\item If a regularization scheme defines $I_\mu=0$
and imposes the trace condition $g^{\mu\nu}I_{\mu\nu}=0$, this
actually yields $I_{\mu\nu}=0$ due to the fact $I_{\mu\nu}\propto
g_{\mu\nu}$, then the $b$-linear parts of $\Gamma^{(1)}$ and
$\Gamma^{(2)}$ will yield the results obtained by Jackiw et al and
Chan et al, respectively,
\begin{eqnarray}
\Gamma^{(1)}_{\rm CS}&=&\frac{1}{16\pi^2}b_\mu\int
d^4x\,\epsilon^{\mu\nu\lambda\rho}F_{\nu\lambda}A_\rho ;
~~~\Gamma^{(2)}_{\rm CS}=\frac{3}{32\pi^2}b_\mu\int d^4x\,
\epsilon^{\mu\nu\lambda\rho}F_{\nu\lambda}A_\rho.
\end{eqnarray}
\item In a regularization scheme defining $I_\mu=0$ and
$I_{\mu\nu}=1/2 g_{\mu\nu}$ (i.e. omitting the surface term of
$I_{\mu\nu}$), the conclusion argued by Coleman and Glashow will
be reproduced,
\begin{eqnarray}
\Gamma^{(1)}_{\rm CS}=\Gamma^{(2)}_{\rm CS}=0.
\end{eqnarray}
\item If defining the one-loop quantum effective action as an
arbitrary combination of (\ref{form1}) and (\ref{form2}) and
choosing a regularization scheme imposing $I_\mu=I_{\mu\nu}=0$,
one can get the result obtained by Chen in differential
regularization,
\begin{eqnarray}
\Gamma_{\rm CS}=(1-c)\Gamma^{(1)}_{\rm CS}+c\Gamma^{(2)}_{\rm CS}
=\frac{2+c}{32\pi^2} b_\mu \int d^4x\,\epsilon_{\mu\nu\lambda\rho}
F_{\nu\lambda}A_{\rho} {\equiv}C b_\mu \int
d^4x\,\epsilon_{\mu\nu\lambda\rho} F_{\nu\lambda}A_{\rho}.
\end{eqnarray}
\end{itemize}
The above explanation shows that the origin for the finite
ambiguous CS term lies in the inequivalence between two
formulations (\ref{form1}) and (\ref{form2}) of the quantum
effective action (\ref{eq:efac}) and the ambiguity of a
regularization method in manipulating logrithmically and linearly
divergent loop momentum integrals.

P\'{e}rez-Victoria\cite{victoria2} put forward another explanation
through revealing a relation between the radiatively induced CS
coefficient and triangle chiral anomaly via an intermediate model
having spontaneous breaking of Lorentz and CPT symmetries. The
fermionic sector of this model takes following
form\cite{victoria2},
\begin{eqnarray}
{\cal L}^{\prime}_{\rm
fermion}=\overline{\psi}\left(iD\hspace{-2.4mm}/\,
-m+\frac{\widetilde{b}}{\Lambda}\gamma_5\partial\hspace{-2mm}/\,\phi
+ic\gamma_5\phi+\frac{d}{\Lambda}\phi^2\right)\psi, \label{model2}
\end{eqnarray}
where $\phi$ is a real pseudoscalar field (i.e. an axion),
$\Lambda$ is certain large scale and $\widetilde{b}$, $c$ and $d$
are indefinite parameters. In the choice that
\begin{eqnarray}
c=d=0,~~ \langle\phi\rangle=\frac{\Lambda}{\widetilde{b}}\,b_\mu
x^\mu, \label{parameter}
\end{eqnarray}
the above model will restore the fermionic Lagrangian
(\ref{fermion1}).

As initially argued by Coleman and Glashow\cite{cogl}, the
radiatively induced CS coefficient $\widetilde{k}$ in the model
(\ref{model2}) can be detected by evaluating the quantum vertex
$\Gamma_{\mu\nu} (p,q)$ composed of one axion and two
photons\cite{victoria2}, i.e., the 1PI part of the correlation
function $\langle A_\mu (p) A_\nu (q)\phi (-p-q)\rangle$,
\begin{eqnarray}
\Gamma^{\mu\nu}(p,q)=\epsilon^{\mu\nu\lambda\rho}p_\lambda q_\rho
C(p,q),~~~\widetilde{k}=-\frac{\Lambda}{2}C(0,0). \label{kb}
\label{1pi}
\end{eqnarray}
The Lagrangian (\ref{model2}) shows that the
$\widetilde{b}$-linear part of $\Gamma_{\mu\nu}(p,q)$ is
explicitly related to the chiral triangle amplitude
$V^{\mu\nu\rho}(p,q)=\langle j^\mu (p)j^\nu (q)j^\rho_5
(-p-q)\rangle$,
\begin{eqnarray}
\Gamma_{\widetilde{b}}^{\mu\nu}(p,q)=\frac{\widetilde{b}}{\Lambda}
e^2(p_\rho+q_\rho)V^{\mu\nu\rho}(p,q). \label{bl1pi}
\end{eqnarray}
It is well known that $V^{\mu\nu\rho}$ satisfies the celebrated
anomalous Ward identity,
\begin{eqnarray}
(p_\rho+q_\rho)
V^{\mu\nu\rho}(p,q)=2imV^{\mu\nu}(p,q)+\epsilon^{\mu\nu\lambda\rho}
p_\lambda q_\rho {\cal A}, \label{aw}
\end{eqnarray}
where ${\cal A}$ is the chiral  anomaly coefficient. Further, the
tensor structure of the canonical term $V^{\mu\nu}(p,q)$, which
comes from an explicit breaking of chiral symmetry by fermionic
mass term, takes the following form,
\begin{eqnarray}
V^{\mu\nu}(p,q)=\langle j^\mu (p) j^\nu (q) j_5 (-p-q)\rangle=
\epsilon^{\mu\nu\lambda\rho}p_\lambda q_\rho V(p,q),
~~~j_5{\equiv}\overline{\psi}\gamma_5\psi, \label{cf}
\end{eqnarray}
Eqs\,(\ref{1pi})-(\ref{cf}) establish a relation among the form
factor $C_{\widetilde{b}}(p,q)$ of the $\widetilde{b}$-linear
sector of $\Gamma^{\mu\nu}(p,q)$, the form factor $V(p,q)$ of the
canonical term in the anomalous Ward identity (\ref{aw}) and  the
anomaly coefficient\cite{victoria2},
\begin{eqnarray}
C_{\widetilde{b}}=\frac{\widetilde{b}}{\Lambda} e^2\left[2m
V(p,q)+{\cal A}\right].\label{rela}
\end{eqnarray}
The canonical term $V^{\mu\nu}(p,q)$ in the anomalous Ward
(\ref{aw}) identity is finite and unambiguous. A comparison
$\langle j^\mu (p) j^\nu (q) j_5 (-p-q) \rangle$ with
$\Gamma_{\mu\nu}(p,q)$ shows that $V^{\mu\nu}(p,q)$ is actually
equal to the $c$-linear part of $\Gamma^{\mu\nu}(p,q)$. It turned
out that the $c$-linear part of $\Gamma^{\mu\nu}(p,q)$ could be
easily calculated since it is convergent\cite{victoria2},
\begin{eqnarray}
\widetilde{k}_c=
-\frac{ce^2\Lambda}{2}V(0,0)=\frac{ce^2\Lambda}{8\pi^2 m}.
\label{kc}
\end{eqnarray}
An insertion of Eqs.\,(\ref{kb}) and (\ref{kc}) into (\ref{rela})
immediately lead to a relation among $k_{\widetilde{b}}$, i.e. the
coefficient of $\widetilde{b}$-linear part of
$\Gamma^{\mu\nu}(p,q)$ defined on the mass-shell of a photon, the
parameter $\widetilde{b}$ and anomaly coefficient ${\cal A}$,
\begin{eqnarray}
\widetilde{k}_{\widetilde{b}}=e^2\widetilde{b}\left(\frac{1}{4\pi^2}
-\frac{1}{2}{\cal A}\right).
\end{eqnarray}
Upon choosing the parameters shown in (\ref{parameter}), one can
find a relation between radiatively induced CS coefficient of the
model (\ref{fermion1}) and chiral anomaly
coefficent\cite{victoria2},
\begin{eqnarray}
k_\mu=e^2 b_\mu\left(\frac{1}{4\pi^2}-\frac{1}{2}{\cal A}\right).
\label{relation}
\end{eqnarray}
The origin of the ambiguity on radiatively induced CS coefficient
is thus revealed since the chiral anomaly coefficient is ambiguous
and regularization dependent. This fact was explicitly and
quantitatively shown in differential
regualrization\cite{djl,hala}.

\section{Summary and Discussion}

 The issue on radiatively induced Lorentz and CPT violation
 in quantum electrodynamics is reviewed and it has not been completely
 settled down yet. In this talk we emphasize how the
 CS term is induced from quantum correction.
 The physical effects it causes in classical electrodynamics  were described
 by Jackiw\cite{jackiw2}.
  Since the present astrophysical observation data has
 excluded the physical consequence of the CS term, thus no matter how
 this $k_\mu$ arises, either as a radiative correction induced from
 the fermionic sector or as free parameter set up by hand, it
 must vanish. Here let us discuss how to ``input" this
 ambiguous quantum correction to compare with astrophysical
 observation. There are two approaches  in the
 literature to include this CS term. One is
 starting from the conventional QED plus an explicit Lorentz and
 CPT violating in the fermionic sector, i.e.
 \begin{eqnarray}
{\cal L}=-\frac{1}{4}F_{\mu\nu}F^{\mu\nu}+\overline{\psi}
\left(i\partial\hspace{-1.7mm}/\,-eA\hspace{-1.8mm}/\,-m\right)\psi
-\overline{\psi}b\hspace{-1.4mm}/\,\gamma_5\psi ={\cal L}_{\rm
QED}-{\cal L}_b, \label{mo1}
\end{eqnarray}
and the CS term (\ref{csterm}) will be induced from quantum
correction with the coefficient $k_\mu \propto e^2 b_\mu$. The
other one is introducing the CS term (\ref{csterm}) at classical
level,
\begin{eqnarray}
\widetilde{\cal L}={\cal L}_{\rm QED}-{\cal L}_b+{\cal L}_{\rm
CS}, \label{mo2}
\end{eqnarray}
and the coefficient of CS term is a free parameter. The {\it
radiatively induced} CS terms calculated in these two models have
different meanings from the viewpoint of renormalization theory,
despite that the processes of calculating CS term and the results
are identical. In the framework described by ${\cal L}$, the
induced CS term can only be considered as a radiative correction,
while in the later model $\widetilde{\cal L}$, depending on the
renormalization condition, the induced CS term can be cancelled by
a finite counterterm and keep the classical parameter $k_\mu$ as
the renormalized one. According to the perturbation theory of a
renormalizable quantum field theory, a quantum correction
calculated in certain regularization scheme, now matter how it is,
finite or infinite, has no physical meaning before a
renormalization procedure is implemented. Only when a
renormalization condition is assigned, the quantum correction is
decomposed into two parts, one part will be cancelled by certain
counterterm and absorbed into the classical Lagrangian to redefine
the various parameters such as mass or coupling constant, the
other part is the radiative correction and reflects the observable
quantum effects. Based on this fact, it can be easily seen that in
the first model the induced CS term cannot be cancelled by
introducing a counterterm since its has no counterpart in the
classical Lagrangian, hence it can only belong to the radiative
correction like chiral anomaly and anomalous magnetic moment etc.
However, the induced CS term, despite of being a radiative
correction, has an essential difference with anomalous magnetic
moment and chiral anomaly: it can not be ambiguously fixed by the
principle of a quantum field theory itself. Anomalous magnetic
moment can be uniquely evaluated by a gauge invariant
regularization scheme. Chiral anomaly, despite it is ambiguous,
can be determined if vector gauge symmetry is required. Whereas
for the induced CS term, it seems that gauge symmetry cannot
dominate it\cite{victoria2}. One may say that the CS term can be
determined by the naturalness of Lorentz and CPT symmetry, but the
naturalness of this space-time symmetry can only take $k_\mu$ to
zero. The ambiguity of the induced CS term make the theory
(\ref{mo1}) awkward, since this means that the theory cannot make
a definite prediction on the quantum phenomena. If we recall the
explanation by Sitenko \cite{sitenko} on the origin of this
ambiguity, this specific example seems to imply an inkling that a
quantum field theory, as the most successful framework of
describing subatomic physics up to now, has an intrinsic
deficiency in certain specific situation such as Lorentz symmetry
breaking\cite{pc}. This speculation might be comprehensible, since
a relativistic quantum field theory was born out of a combination
between quantum theory and the theory of special relativity. If
Lorentz symmetry, the algebraic soul of special relativity,
collapsed, what could one expect from a relativistic quantum field
model! Of course, it may not be  so serious, since there exists a
possible way out for the model (\ref{mo1}), namely, considering
the contribution to the CS coefficient from all the fermion
species in the Standard Model \cite{choh},
\begin{eqnarray}
k_\mu \propto \sum_{i}e_i^2 b_\mu^i,
\end{eqnarray}
where the sum index $i$ runs over all the leptons and quarks of
the Standard Model. If $b_\mu^i$ come from the vacuum expectation
value $\langle A_{\mu}^{5} \rangle$ of an axial vector gauge
field, then the induced CS coefficient may vanish according to the
anomaly cancellation in the Standard Model. However, as indicated
by Chung et al\cite{choh}, it is also possible that $b_\mu$ may
not be related to the vacuum expectation value of an axial vector
field.

  In contrast, the second setting (\ref{mo2}) is more appealing in discussing
  the physical effects of CS term. Since CS term
  is put by hand at the classical level, one can introduce a finite
  counterterm to define renormalized CS
  coefficient\cite{bo,victoria2}
  \begin{eqnarray}
  k^\mu_{\rm ren}=k^{\mu}_{\rm bare}+k_{\rm quant}^\mu+k_{\rm
  counter}^\mu,
  \label{ren1}
  \end{eqnarray}
and then input $ k^\mu_{\rm ren}=0$ to yield to the astrophysical
observation data. However, in this case, both $k^\mu_{\rm ren}$
and $b_\mu$ are regarded as independent measurable parameters,
$k^\mu_{\rm ren}$ has nothing with $b_\mu$ and the astrophysical
observation data about the vanishing  of $k_\mu$ does not put any
constraint on $b_\mu$ \cite{bo,victoria2}. A fine-tuning is
required to get a vanishing $k_\mu$ and non-vanishing $b_\mu$ at
the same time since the CS term can be generated from quantum
correction with the coefficient proportional to
$b_\mu$\cite{victoria2}.

 Finally, a conclusion from an investigation on the anomalous
magnetic moment and Lamb shift in QED with the extended fermionic
sector (\ref{fermion1})  should be emphasized. It was
found\cite{chku} that both the anomalous magnetic moment and Lamb
shift received an additional IR divergent radiative correction
proportional $b^2$. Furthermore, it was explicitly shown that even
the IR divergence in the Lamb shift cannot be cancelled by the
bremsstrahlung process as in the conventional QED\cite{chku}, let
alone eliminating the IR divergence in the anomalous magnetic
moment. Since the anomalous magnetic moment and Lamb shift are two
successful symbols of QED in describing electromagnetic
interaction, the IR divergence embracing them seems thus to
reflect the physical inconsistency of  QED with above extended
fermionic sector, and therefore ruins the mechanism of generating
CS term from radiative correction by introducing an explicit
Lorentz and CPT violating term in the fermionic sector. Of course,
there is a possibility that the calculation on the vertex
correction\cite{chku} has some drawbacks, since we only expanded
the $b$-exact propagators to the second order, perhaps a summation
to any order of $b_\mu$ might erase such an IR divergence.

 \section*{Acknowledgment}
I would like to thank the organizers of PASCOS 2001, Professors P.
Frampton and Y.J. Ng and the organizers of MRST 2001, Professors
V. Elias, D.G.C. McKeon and V.A. Miransky for hospitality. I am
grateful to Professor R. Jackiw for suggesting me this project and
various useful discussions.  I am indebted to Professors M.
Chaichian, G. Kunstatter and Dr. R. Gonz\'{a}lez Felipe for
discussion and collaboration. I would like to thank Professors G.
Bonneau, L.H. Chan, V.A. Kosteleck\'{y}, \fbox{G. Leibbrandt} and
R.B. Mann for communications and discussions. I am especially
obliged to Dr. M. P\'{e}rez-Victoria for his continuous
discussions and comments. This work was supported in part by the
Natural Sciences and Engineering Research Council of Canada.


\begin{thebibliography}{99}
\bibitem{pc}There is a comment from Dr. M. P\'{e}rez-Victoria on this
argument: According to the prescription of renormalization theory,
 the bare Lagrangian should
consist of all the possible renormalizable terms allowed by gauge
symmetry, while ${\cal L}_{\rm CS}$ is just such a term. Thus the
Lagrangian (\ref{mo1}) is actually incomplete and one must start
from the classical Lagragian (\ref{mo2}) to discuss the quantum
correction. This is the reason why the quantum theory based on
(\ref{mo1}) is not well defined. Therefore, there should exist no
problem for a quantum field theory with  the breaking of Lorenz
symmetry if one start from the complete Lagrangian (\ref{mo2}).
The model (\ref{mo1}) can be understood as a special case of
(\ref{mo2}) with the renormalized $k_\mu$ happenning to to zero.
My comprehension is that (\ref{mo1}) and (\ref{mo2}) are just two
different model construction and they describe distinct classical
electrodynamics. The quantum phenomena of these two models thus
have different intepretation, for example, the radatively induced
CS term in the model (\ref{mo1}) can only be explained as
radiative correction, while in (\ref{mo2}) it can be erased by
redefining the CS parameter according to the finite
renormalization procedure. Of course, classical dynamics is an
approximation of the corresponding quantum counterpart. In this
sense, the model (\ref{mo2}) is more appealing  since it has a
well defined qunatum theory. However, in some case, the model
(\ref{mo1}) rather than (\ref{mo2}) can come from a the low-energy
limit of a more fundamental theory.
\end{thebibliography}
\end{document}